Compensation strategies for a lip-tube perturbation of French [u]:

an acoustic and perceptual study of 4-year-old children

Lucie Ménard[a], Pascal Perrier[b], Jérôme Aubin[a], Christophe Savariaux[b] and Mélanie Thibeault[a]

[a] Laboratoire de Phonétique, Département de linguistique et de didactique des langues,

Université du Québec à Montréal,

CP 8888, Succ. Centre-Ville, Montréal, H3C 3P8, Canada

[b] ICP-INPG, UMR CNRS No. 5009, Université Stendhal,

Boîte Postale 25, 38040 Grenoble Cedex 9, France

Suggested running head: Compensation for a lip-tube perturbation

## ABSTRACT

The relations between production and perception in 4-year-old children were examined in a study of compensation strategies for a lip-tube perturbation. Acoustic and perceptual analyses of the rounded vowel [u] produced by twelve 4-year-old French speakers were conducted under two conditions: normal and with a 15-mm-diameter tube inserted between the lips. Recordings of isolated vowels were made in the normal condition before any perturbation (N1), immediately upon insertion of the tube and for the next 19 trials in this perturbed condition, with (P2) or without articulatory instructions (P1), and in the normal condition after the perturbed trials (N2). The results of the acoustic analyses reveal speaker-dependent alterations of F1, F2, and/or F0 in the perturbed conditions and after the removal of the tube. For some subjects, the presence of the tube resulted in very little change; for others, an increase in F2 was observed in P1, which was generally reduced in some of the 20 repetitions, but not systematically and not continuously. The use of articulatory instructions provided in the P2 condition was detrimental to the achievement of a good acoustic target. Perceptual data are used to determine optimal combinations of F0, F1, and F2 (in Bark) related to these patterns. The data are compared to a previous study conducted with adults [Savariaux *et al.*, J. Acoust. Soc. Am. **106**, 381–393 (1999)].

## I. INTRODUCTION

In recent decades, artificial perturbation of the speech articulators has proven to be a very fruitful experimental paradigm. Indeed, substantial research conducted within this framework has shed light on the nature of the internal representations of vowels and consonants and the role of feedback (auditory, proprioceptive, etc.) in controlling the vocal apparatus (Aasland *et al.*, 2006; Jones and Munhall, 2003; Houde and Jordan, 1998; Guenther *et al.,* 1998; McFarland *et al.*, 1996; Savariaux *et al.*, 1995).

With regard specifically to compensatory abilities in children, perturbation experiments have led to somewhat contradictory results. For instance, in a bite-block experiment conducted on a 4-year-old and an 8-year-old subject, Oller and MacNeilage (1983) concluded that children cannot achieve complete compensation in the spectral domain. Speakers were instructed to repeat productions of /i/ and /æ/ in both free-mandible and fixed-mandible conditions (with a bite-block). Spectrographic analysis revealed differences between the two conditions, suggesting that the children did not fully compensate for the perturbation, in the spectral domain. However, listeners judged the stimuli to be fairly good, which led the researchers to conclude that compensatory strategies could consist in preserving other acoustic parameters (duration, for instance). These results partly confirm Gibson and McPhearson's (1980) study. Those authors instructed 6- and 7-year-old subjects to produce Swedish vowels with and without bite-blocks inserted. Acoustic measurements showed partial compensations for the perturbation in the spectral domain, a pattern which was confirmed by perceptual assessment (vowels were less accurately transcribed). However, a later articulatory study suggested that 4-year-old children exhibit adult-like compensatory abilities. Smith and McLean-Muse (1987) studied lip and jaw displacement and velocity in vowels produced in normal and bite-block conditions by three groups of subjects: 4- and 5-year-old children, 7- and 8-year-old children, and adults. Although the children produced more within-speaker articulatory variability than the adults, all three subject groups showed comparable compensatory abilities. Thus, the authors concluded that this ability is acquired early in childhood and requires very limited language experience. Similar results were obtained by Baum and Katz (1988), in a study of five speakers in each of the following age groups: 4- to 5-year-olds and 7- to 8-year-olds. Speakers were instructed to repeat the vowels [i a u] in normal and bite-block conditions. Acoustic measurements of the first two formant frequencies extracted at vowel onset and vowel midpoint revealed no significant differences between perturbed and unperturbed trials, for both groups of children. Furthermore, speakers compensated at vowel onset in the perturbed trials, suggesting that no auditory feedback was required in this process. Campbell (1999) also suggests that children do not rely on auditory feedback to produce compensatory articulatory strategies in speech.

Together, these studies suggest conflicting conclusions regarding 4-year-old children's compensatory abilities when they are instructed to produce vowels while the jaw is fixed by a bite-block. This type of perturbation does not modify the geometry of the vocal tract: it keeps one articulator from contributing to the constriction area and location typical of this vowel category, but without perturbing the geometry. Another type of perturbation consists of a lip-tube inserted between the lips. This perturbation not only prevents the speaker from closing the lips while producing the vowel [u], for instance, but it also forces a complete reorganization of the articulatory and geometric strategies used to perform the speech task. This kind of perturbation was studied by Savariaux *et al.* (1995) in an articulatory and acoustic experiment conducted on 11 speakers producing the French vowel [u] in the normal condition and with a 2-cm-diameter lip-tube inserted between the lips (perturbed condition). During the first perturbed trial, upon insertion of the lip-tube, seven speakers moved their tongues backwards, presumably to limit the expected deterioration of the acoustic signal. However, none of them achieved full compensation in this trial. In the remaining 19 perturbed trials, those seven speakers used auditory feedback to develop compensation strategies (backward movement of the tongue to shift the constriction location from the velo-palatal to the velo-pharyngeal region). One speaker showed complete compensation in the F1/F2 domain, while four speakers did not compensate at all. The authors proposed that the great between-speaker variability in the compensation strategies reveals the speaker-specificity of the internal representation of the articulatory-acoustic relations in the region of the French vowel [u].

A perceptual study of the vowels produced in normal and perturbed conditions (Savariaux *et al.*, 1999) revealed that globally, instances of /u/ produced in perturbed conditions were less intelligible than those produced in the normal condition. Perceived vowels were represented in an acoustic-auditory space defined by linear combinations of F1, F2, and F0, in Bark. A reinterpretation of the acoustic data in light of the perceptual space revealed that speakers who produced compensatory maneuvers altered F1, F2, and F0 in order to achieve an acoustic target. Thus, speakers had a good knowledge of the articulatory-to-acoustic relationships related to /u/.

In a follow-up study, Savariaux *et al.* (1997) performed an acoustic experiment aimed at examining the exact nature of F0 alterations in speakers and the role played by articulatory guidance. Subjects were instructed to produce /ogu/ sequences in normal and perturbed conditions (with the lip-tube inserted between the lips). The results showed that three subjects achieved complete compensation in this condition. Thus, articulatory instructions inducing a general posterior positioning of the tongue, which is close to the appropriate tongue shape to compensate for the lip-tube perturbation, improved the speakers' ability to recalibrate their articulatory-to-acoustic maps. To our knowledge, no such study has been conducted with child speakers.

The present experiment was designed to extend Savariaux's (1995; 1997; 1999) studies and investigate abilities to compensate for a lip-tube perturbation in 4-year-old French children. If children already exhibit adult-like compensation strategies, it might be suggested that motor equivalence and internal representations of the speech apparatus are acquired early and do not require extended language experience. On the other hand, if children achieve only limited compensation for the perturbation, it would appear that more elaborate internal representations and motor control are required, together with extended language experience, to produce flexible articulatory strategies. The issue of the role of auditory feedback in formulating compensation strategies will also be addressed by comparing acoustic measurements extracted immediately upon insertion of the lip-tube to measurements extracted in subsequent perturbed trials. Since the experimental design and data analysis method are similar to those of Savariaux's studies conducted with French adults (Savariaux *et al.*, 1995; 1999), results of the present study will be compared to those obtained with the adult group.

## II. EXPERIMENT 1

### A. Method

*1. Subjects*

Twelve French children ranging in age from 3 years 10 months to 4 years 11 months participated in this study. Speakers were recruited in a day-care center and were all native speakers of Continental French. The children were monolingual speakers of French living in the Grenoble (French Alps) area. They had no history of speech or language difficulties (as determined by the day-care center's professionals). Children were also given a pretest screening by two of the experimenters in order to ensure that they had no oral cavity anomalies.

*2. Material and procedure*

Acoustic recordings of the French vowel [u] were made using a digital audio tape recorder (TASCAM) and a high-quality microphone. Subjects were instructed to repeat the isolated vowel [u] in four conditions: normal condition before lip-tube insertion (hereafter N1), with the lip-tube inserted between the lips (hereafter P1), with the lip-tube inserted between the lips and the instructions to start from [o] (hereafter P2), and in the normal condition after removal of the lip-tube (N2). The P2 condition was added following Savariaux *et al.* (1997), who showed that this condition provided articulatory instructions that improved the subject's compensatory response. The instructions were as similar as possible to those provided to adult subjects in the Savariaux *et al*.'s (1995; 1997) studies. Throughout the perturbed phrases, the experimenters reminded the subjects that the target was the vowel /u/. The order of the two perturbed conditions was counterbalanced across subjects. Subjects produced 20 repetitions of [u] in each condition. For one subject, 12 repetitions were produced in P2 and 15 trials were recorded in P1. Six repetitions of each of the vowels [i a o ɔ œ] in normal condition were also recorded.

A 1.5-cm-long lip-tube was built of Plexiglas. The length was chosen so as to avoid lengthening the labial constriction. The diameter was chosen using simulations with an articulatory-to-acoustic model integrating non-uniform vocal tract growth (Variable Linear Articulatory Model, hereafter VLAM) (Boë and Maeda, 1999). This model is a scaled version of an adult model (Maeda, 1989), based on Goldstein's (1980) anatomical data from birth to adulthood. This model generates realistic vocal-tract shapes and has been used by our group in various studies (Ménard *et al.,* 2002; Ménard *et al.,* 2004; Ménard *et al,* 2007). The model is controlled by seven articulators, which represent functional articulatory blocks: lip (height and protrusion), jaw height, tongue position (tip, body, back), and larynx height. The model provides a sagittal view of the vocal tract shape, the corresponding area function (based on Heinz and Stevens, 1955), and the transfer function (Badin and Fant, 1984). VLAM allowed us to determine the optimal diameter for the lip-tube. The model was set at the 4-year-old stage, and the lip height parameter was increased so that the lip area value increased as well. A resulting value of 1.77 cm$^2$ was chosen, which corresponds to a 1.5-cm diameter. Figure 1 shows, as predicted by the model, the percentage variation in F1 and F2 resulting from the perturbation, relative to the prototypical [u] for that stage (N1). This condition, representing the formant values with the lip-tube inserted but without any compensatory maneuver, will be referred to as "Pert". Figure 1 reveals that the insertion of the lip-tube, without any compensatory maneuvers, results in a 42% increase in F1 and a 32% increase in F2. Those simulated changes are in the range of those reported in Savariaux *et al.* (1995; 1999) and the perturbation can thus be considered equivalent, considering the children's smaller vocal tract. Because /u/'s F1 and F2 are affiliated to Helmholtz resonators, those formant frequencies are sensitive to various changes in constriction area and constriction length. Indeed, a Helmholtz's resonance frequency can be calculated by the following formula: $F = (c/2\pi k)\sqrt{(A_{co}/L_{co}V_{ca})}$ where $c$ is sound velocity, $k$ is a factor accounting for variation in the cross-section shape, $A_{co}$ is constriction area, $L_{co}$ is constriction length, and $V_{ca}$ is cavity volume. In order to decrease formant values, several strategies can be used, possibly in combination: decrease constriction area, increase constriction length or increase cavity volume. Local changes can be made in any of these dimensions, which results in small alterations in formant values. However, simulations carried out with VLAM suggest that the best compensatory strategy resulting in minimal formant alterations compared to the non perturbed condition N1 (labelled "Pert-comp" in Figure 1) involves a backward and downward movement of the tongue body, which increases constriction length and reduces constriction area (because of the shape of the palate, the downward movement of the tongue is necessary to avoid complete occlusion). Those strategies decrease the frequency of the affiliated formant and thus counterbalance the increase in frequency related to the decrease in cavity volume of the back cavity. This displacement results in full compensation for the perturbation. The articulatory compensation strategy (Pert-comp) lowers F1 and F2 so that they almost reach the N1 value.

[Insert Fig. 1]

*3. Data analysis*

All vowels were digitized at a rate of 44100 Hz with 16-bit quantization. Data were downsampled at 22050 Hz after low-pass filtering at a cut-off frequency of 11025 Hz, in order to obtain a more accurate formant detection in the [0, 44000Hz] range. The first two formant frequencies were then extracted for each vowel, using the LPC algorithm integrated in the Praat speech analysis program, at vowel onset (first glottal pulse) and at vowel midpoint. The number of poles varied from 10 to 14 (number of LPC coefficients from 20 to 28), which is in the range of parameters used by Lee *et al.* (1999) and Hillenbrand *et al.* (1995). A 14-ms Hamming window was used, with a pre-emphasis factor of 0.98 (pre-emphasis from 50 Hz for a sampling frequency of 22050 Hz). It is well known that formant measurements are particularly difficult to extract in high-pitched voices, due to the large distance between adjacent harmonics, leading to undersampled spectra. This is especially important for LPC analyses, in which formant measures are greatly influenced by the closest harmonic (Atal and Schroeder, 1974). However, LPC analysis is the procedure used by recent acoustic studies of child speech (Lee *et al.*, 1999; Hillenbrand *et al.*, 1995). Thus, we tried to avoid formant measurement errors by comparing, for each vowel, the automatically extracted formant values overlaid on a wide-band spectrogram with a spectral slice obtained by an FFT computation with a Hamming window. When large discrepancies were observed either (i) between the overlaid formant values and the spectrogram or (ii) between the overlaid formant values and the spectral slice, the prediction order of the automatic detection algorithm was readjusted and the analysis was performed again.

*4. Statistical analysis*

For the production part of the study, three separate repeated-measure ANOVAs were carried out with experimental condition (N1, P1, P2, and N2) and measurement point (vowel onset, vowel midpoint) as the within-subject factors, using F1, F2, or F0 as the dependent variable. Data from the 12 subjects were included in the analysis. Another set of three repeated-measure ANOVAs was conducted on spectral measurements for the first and last trials only in each of the four conditions. Thus, the independent variables were experimental condition (N1, P1, P2, and N2) and trial number (first and last). For each analysis, interaction effects were further explored by planned comparisons with the alpha level set to 0.05[i]. This design was chosen to reveal global trends in speech production data, following Max and Onghena (1999). However, because between-speaker variability is often reported in such studies, individual behaviors will also be described. It should be noted that the order of the perturbed conditions (P1 followed by P2 or P2 followed by P1) did not significantly influence the data produced in each condition. Indeed, the results of T-tests carried out on F1, F2, and F0 values did not reveal a significant difference between the data for the six subjects who performed P1 before P2 and the data for the remaining six subjects who performed P2 before P1. Thus, in the following analyses, the order of the conditions will be presented but this factor was not included in the design of the ANOVAs.

**B. Results and discussion**

Mean F1 and F2 values for each subject and condition, at each measurement time (first glottal pulse or midpoint) are shown in Tables 1 and 2. The measurements extracted at the vowel midpoint will be presented first, since we considered this data point to be representative of the vowel's target value. The measurement extracted at the vowel onset will be presented in order to evaluate whether compensation occurred immediately. Standard deviation values are presented in square brackets. For the sake of clarity, the percentage variation relative to the normal pre-perturbed condition (N1) is also presented. For each speaker, the order of elicitation of the perturbed conditions P1 and P2 is shown in subscript. Recall, however, that since no significant effect of this factor on the formant and F0 values was found, data in each of the perturbed conditions were pooled together and the elicitation order was ignored in the analysis. Graphic representations of the mean values and the standard deviations of the mean F1 and F2 values are provided in Figure 2.

[Insert Table 1 and Figure 2]

*1. Mean spectral measures across conditions*

*a. F1 values.* The data presented in Table 1, for the vowel midpoint in the P1 and P2 conditions, display considerable between-speaker variability. Some speakers produced the [u] vowel in the P1 condition with a minimal increase in F1 relative to N1 (such as S1,S5 and S11), whereas some others produced a large F1 increase in P1 relative to N1 (such as S2 and S8). When provided with articulatory cues (P2 condition), most subjects did not produce F1 values that were any closer to the values in the N1 condition. Thus, the difference in F1 values between the P2 and N1 conditions was no smaller than the difference in F1 between the P1 and N1 conditions; in fact, it was greater. An examination of the onset and the midpoint measurement values suggests a slight decrease in F1 throughout vowel duration for most speakers. In order to assess the effects of condition (N1, P1, P2, N2) and measurement point (vowel onset or midpoint), a two-way repeated-measure ANOVA was conducted on F1 values. The results revealed a significant main effect of experimental condition on F1 values ($F(3,33) = 33.38$; $p < .01$). Post hoc tests revealed that F1 values were significantly higher in both perturbed conditions than in the normal pre-perturbed condition N1 ($F(1,11) = 12.45$; $p < .01$ for P1 and $F(1,11) = 42.23$; $p < .01$ for P2). Values produced in the P2 condition were significantly higher than those produced in the P1 condition ($F(1,11) = 6.00$; $p < .05$). In the post-perturbed condition N2, F1 was significantly lower than in the normal pre-perturbed condition N1 ($F(1,11) = 32.80$; $p < .01$) and in both perturbed conditions: P1 ($F(1,11) = 29.96$; $p < .01$) and P2 ($F(1,11) = 73.51$; $p < .01$), suggesting the existence of a robust after-effect.

Concerning the effects of vowel measurement point, the ANOVA revealed that F1 values were significantly higher at the vowel onset than the vowel midpoint ($F(1,11) = 25.52$; $p < .01$). However, no significant effect of the interaction between measurement point and condition was observed, suggesting that F1 variations throughout the duration of the vowel do not arise from an adjustment in response to the perturbation.

[Insert Table 2]

*b. F2 values.* For F2 values (Table 2 and Figure 2 (right panel)), as was the case for F1, the measurements display significant between-speaker variability. In the P1 condition, at the vowel midpoint, all subjects but one (S7) produced the vowel [u] with an increase in mean F2 value relative to the N1 condition. However, for four subjects (S1, S4, S5, and S12), the mean percentage increase in F2 value is less than 10%. Turning now to the variation in F2 in the N2 condition, Table 2 shows that all subjects but one (S6) produced lower F2 values in the normal post-perturbed condition N2, compared to the pre-perturbed condition N1, confirming a robust after-effect, as was the case for F1 values (Table 1 and Figure 2 (left panel)). However, the mean percentage decrease in F2 values ranges from –36% to –3%, suggesting variability across speakers.

A repeated-measure ANOVA with measurement point (vowel onset and vowel midpoint) and experimental condition (N1, P1, P2, N2) as the within-subject factors was carried out on the 12 subjects' mean F2 values. As shown in Figure 2 (right panel), a significant effect of measurement point was observed ($F(1,11) = 51.32$; $p < .05$). Indeed, F2 was higher at vowel onset than at vowel midpoint. This difference is found in both perturbed conditions and in both normal conditions, as revealed by the lack of a significant interaction effect between measurement point and experimental condition. The effect of the condition factor was significant ($F(3,33) = 48.78$; $p < .05$). Post hoc tests showed that F2 values in N1 were significantly lower than in P1 ($F(1,11) = 32.99$; $p < .01$) and in P2 ($F(1,11) = 22.81$; $p < .01$). F2 values observed in P1 did not differ from those measured in P2. Values in the N2 condition were significantly lower than values in N1 ($F(1,11) = 22.69$; $p < .01$), suggesting a robust after-effect, and they were lower than in both perturbed conditions P1 ($F(1,11) = 93.47$; $p < .01$) and P2 ($F(1,11) = 101.13$; $p < .01$).

*c. F0 values*. Mean F0 values are presented in Table 3 for each subject and each condition. Standard deviations are provided in square brackets and mean percentage increases in the P1, P2, and N2 conditions relative to the N1 condition are shown in parentheses. It is noticeable in Table 3 that the evolution of F0 values over the four experimental conditions varies among the 12 subjects. A repeated-measure ANOVA was computed on the mean F0 values for the 12 subjects with measurement point (vowel onset and vowel midpoint) and experimental condition as the within-subject factors. No significant effects of the factors, either as main effects or in interaction, were observed.

[Insert Table 3]

To summarize, an analysis of the trials produced in all conditions reveals that, overall, speakers were significantly affected by the lip-tube. Indeed, F1 and F2 values were higher in the P1 and P2 conditions compared to N1. In the normal post-perturbed condition N2, formant values were significantly lower than in the normal pre-perturbed condition N1. Considerable between-speaker variability was observed in the extent to which those acoustic parameters were affected by the perturbation.

*2. Learning effects over the perturbed trials*

In order to evaluate the variation in formant values within a given condition across the 20 trials, Figure 3 presents F1 (left panel) and F2 values (right panel) for the first and last trials in the four experimental conditions. Values measured at vowel midpoint are considered here. Mean values and standard deviations are represented for the 12 subjects. The solid line corresponds to the first trial, whereas the dashed line depicts the last trial. Two separate repeated-measure ANOVAs carried out on F1 and F2 values with trial number (first and last) and experimental condition (N1, P1, P2, N2) as the within-subject factors did not reveal any significant effect on the variation in formant values. Thus, overall, speakers did not show any tendency to alter their formant values from the first to the last trials in the F1 and F2 dimensions separately, suggesting that there was no learning effect. Furthermore, no significant within-subject correlations were found between trial number (from 1 to 20) and any of the acoustic parameters. It must be remembered, however, that those data include all subjects. As was observed in previous perturbation experiments (Savariaux *et al.*, 1995, for instance), subjects vary in the extent to which they respond to the perturbation.

In summary, in spite of considerable inter-speaker variability, clear trends emerge from the analysis of the F1, F2, and F0 values produced by the subjects in normal and perturbed conditions: for the majority of the subjects, F1 and F2 are significantly modified by the insertion of the lip-tube in the P1 and P2 conditions; none of the subjects systematically improved his/her (F1, F2) patterns during the perturbed phase in either the P1 condition or the P2 condition. These two observations suggest that, in general, subjects were not able to develop a robust compensation strategy within the 20 trials of the perturbed phase. However, a robust after-effect is observed for all the subjects, suggesting the possibility that they may have learned new articulatory-to-auditory internal representations. Hence, speakers may have demonstrated abilities in the acoustic-auditory space, defined by linear combinations of spectral measures (Ménard *et al.*, 2002; Savariaux *et al.*, 1999). The following section presents a perceptual description of the stimuli produced by the subjects, to determine whether strategies to compensate for the presence of the lip-tube were successful.

## III. EXPERIMENT 2

Previous experiments have shown that invariant correlates of perceived vowels can be found in linear combinations of spectral parameters. The identification of openness, for instance, has been found to be related to the difference between F0 and F1 in various languages (Traunmüller, 1981; Syrdal and Gopal, 1986; Hoemeke and Diehl, 1994; Ménard et al., 2002). Prior to the computation of this difference, Hertz values are converted into a semi-logarithmic scale, usually the "critical band units" (Bark) scale, based on psychoacoustic experiments (Potter and Steinberg, 1950; Traunmüller, 1981). In French, high vowels (like /u/) would be related to a distance between F1 and F0 of less than 2 Bark (Ménard *et al.*, 2002). Concerning place of articulation, the F3–F2 difference would account for the perception of this feature in English (Syrdal and Gopal, 1986), whereas F2–F1 would be involved for the same feature in Swedish (Fant, 1983). Hirahara and Kato (1992) propose that F2–F0 is related to the perception of place of articulation. Similarly, Savariaux *et al.* (1999) found that the perception of /u/ in French was associated with a low value for the F2–F0 difference and/or with a low F1 value. Thus, for /u/, combinations of formant frequencies and F0 appear to be involved in the perceptual identification of the speech target. The goal of this second experiment is twofold. First, the results will be used to determine to what extent subjects were able to achieve a good compensation strategy to produce the perceptual target related to /u/. Second, the acoustic parameters related to the perceptual identification of /u/ will be determined and used to reinterpret the production data presented in Section I.

### A. Method

*1. Subjects and stimuli*

A subset of the produced vowels were used as stimuli for a perceptual test, as in Savariaux *et al.* (1999) and McFarland *et al.* (1996). For each speaker, the first five vowels produced in the normal pre-perturbed (N1) and post-perturbed condition (N2) and 10 vowels in each of the perturbed conditions (P1 and P2) were selected. The selected vowels in the perturbed conditions were those from the first and last trials, as well as the odd-numbered trials 3, 5, 7, 9, 13, 15, 17, and 19. This selection was representative of the whole set of stimuli produced in each condition, while keeping the number of stimuli reasonably low. Separate T-tests with F0, F1, and F2 as the

Compensation for a lip-tube perturbation 17variables performed between the 10 selected vowels for the perceptual test and the remaining 10 vowels produced in the first experiment (N1) by each speaker did not reveal any significant differences between the two datasets. One repetition of each of the vowels [i a o] was also included. As a result, a total of 396 vowels (33 vowels*12 speakers) constituted the set of stimuli. The total duration of the test was 40 minutes. Fifteen native speakers of French, ranging in age from 22 to 35 years old, served as subjects for the experiment. The participants did not report any history of auditory abnormality or speech production disorder. The test took place in a quiet room, on a Toshiba portable computer, using the perceptual experiment procedure implemented in Praat. Vowels were presented once binaurally via high-quality headphones. The tests consisted of an identification task and a quality-rating task. Participants had to (i) identify the vowel they heard from among the French oral vowels /i y u a o œ ɔ/, and (ii) rate the quality of the vowel they heard, if the vowel was /u/. For the latter task, seven choices were available: excellent /u/, very good /u/, good /u/, average /u/, bad /u/, very bad /u/, not a /u/. The participants had to select an icon displayed on the computer screen using the mouse.

*2. Data analysis*

In a first analysis of the perceptual responses, global identification scores were calculated. This parameter corresponds to the percentage of /u/ vowels correctly identified by the listeners. Then, in a subsequent analysis, only stimuli which were perceived as /u/ by at least 50% of the listeners were included in the analysis. These stimuli will be referred to as the dominantly perceived vowels.

*3. Statistical analysis*

For the identification task of the perceptual test, identification scores were computed by dividing, for each vowel produced, the number of responses from the 15 listeners for which the perceived vowel corresponded to the produced vowel. For quality-rating responses, among the vowels perceived as /u/, three categories were considered, based on the average quality rating task: vowels rated "very bad" or "bad" were pooled into one category (referred to as "bad"), vowels rated "average" were labeled "average," and vowels rated "good," "very good" or "excellent" were pooled in a third category labeled "good." Vowels rated "not an /u/" constituted a fourth category. Two separate repeated-measure ANOVAs were carried out on identification scores and mean goodness-rating scores with experimental condition as the within-subject factor (N1, P1, P2, and N2).

### B. Results

*1. Mean identification scores*

[Insert Table 4 and Figure 4]

The results of the perceptual test were used to determine to what extent subjects were able to achieve a good compensation strategy to produce the acoustic-auditory target related to /u/. Table 4 presents the mean percentage of produced /u/ vowels perceived as /u/ in each condition, for each subject. In the P1, P2, and N2 conditions, the percentage variation relative to the N1 condition is presented in parentheses. Mean perceptual scores across the four experimental conditions are depicted in Figure 4. First, Table 4 clearly shows that, even in the N1 condition, the perceptual scores vary among speakers. Indeed, some speakers (such as S1 and S5) produced

/u/ that were generally correctly identified by listeners (over 90% correct), whereas some others (such as S4 and S11) produced /u/ that were associated with low scores (47% and 43%).

A repeated-measure ANOVA was carried out on those values with condition (N1, P1, P2, N2) as the within-subject factor. A significant effect of condition was found ($F(3,11) = 37.88$; $p < .05$). Planned comparisons revealed that the percentage of correct identification was higher in the normal pre-perturbed condition N1 than in the two perturbed conditions, P1 and P2 ($F(1,11) = 37.88$; $p < .05$). Identification scores in N1 were in turn lower than in the normal post-perturbed condition N2 ($F(1,11) = 14.85$; $p < .05$), confirming the robust after-effect found in the F1 vs. F2 domain, and corresponding to a more canonical production than in the N2 condition. Furthermore, scores in P2 were lower than in P1 ($F(1,11) = 13.25$; $p < .05$), suggesting that the use of articulatory instructions was actually detrimental to the achievement of a good acoustic-auditory target. On the basis of those perceptual criteria, then, compensation was better in the P1 condition than in the P2 condition, which was not observed for the F1, F2, and F0 dimensions independently.

## 2. Learning effects over the perturbed trials

In order to determine the possible learning strategies applied in the development of compensatory maneuvers over the course of the 20 trials, the identification of the first and last trials in P1 and P2 was examined. Data are plotted in Figure 5 (P1 in left panel and P2 in right panel). Since considerable between-speaker variability is observed, the identification scores for each of the 12 subjects are presented separately in the graphs. Data corresponding to the first trial are depicted by the solid line, whereas values corresponding to the last trial are represented by the dashed line. The mean identification scores for the N1 condition correspond to the bars.

[Insert Figure 5]

.

To evaluate to what extent the variation in identification from the first to the last trials reflects learning mechanisms, simple regression analyses were carried out between trial numbers and identification scores in P1 and P2, for each subject. No significant correlation was found. The variability observed for each speaker within the perturbed conditions might reflect the incapacity of the children to reproduce identically a vocal tract configuration, rather than their attempt to improve their production. However, as can be noticed in Tables 1 and 2, the standard deviations measured for the acoustic parameters are higher in the two perturbed conditions P1 and P2 than in the two normal conditions N1 and N2. Two separate repeated-measure ANOVAs were carried out on F1 and F2, with measurement points and conditions as the within-subject variables. Both ANOVAs revealed a statistical effect of the condition factor on the standard deviations (for F1: $F(3,33)=12.89$, $p < .05$; for F2: $F(3,33)=13.99$, $p < .05$), with higher values in P1 and P2 than in N1 and N2 (for F1: $F(1,11)= 20.25$, $p < .05$; for F2: $F(1,11)=23.71$, $p < .05$). The larger variability observed in both perturbed conditions compared to the normal pre-perturbed and post-perturbed conditions supports the hypothesis of a search for a compensatory strategy to improve the /u/ production in presence of the perturbation. The absence of any evidence for a learning effect suggests that this search is based on a trial-and-error basis, probably guided by auditory feedback.. Furthermore, those perceptual results suggest that compensatory strategies guided by articulatory information (P2) were not successful, since they did not allow the speakers to improve the identification scores of their vowels

*3. Best perceived /u/ in perturbed conditions*

Since our results suggest that learning does not take place during the 20 trials of the training phase in the perturbed conditions (P1 and P2), it can be assumed that the last trial is not necessarily the best one in terms of perceptual efficiency. Hence, seeking out the best-perceived vowel within the trials produced in perturbed conditions could inform one about each subject's inherent articulatory capability to compensate for the perturbation, independently of the capacity to learn and memorize the best strategy. The percentage of produced /u/ correctly identified as /u/ for this best-perceived trial in P1 and P2 is depicted in Figure 6 for each subject. Values in the P1 condition correspond to the solid line, and values in the P2 condition are depicted by the dashed line. For the sake of clarity, the mean identification scores for the normal trials in the N1 condition are also displayed. Figure 6 reveals that, in the P1 condition, all speakers produced at least one vowel associated with a higher or equal identification score than the mean identification score in the N1 condition. If compensation proficiency is evaluated through the ability to produce a vowel in the perturbed condition associated with an identification score equal to or greater than the mean identification score in the normal condition N1, then all subjects were able to achieve complete compensation in at least one stimulus, during the 10 trials. Seven speakers even had identification scores of greater than 80% and can thus be considered to be the best compensators: S1, S5, S6, S8, S10, S11, S12. Hence, it can be concluded that all speakers had the articulatory skills to fully compensate for the perturbation that was introduced with the lip tube chosen for the experiment. The reasons why they did not do it permanently after having done it once do not originate from articulatory skills.

[Insert Figure 6]

As could be expected from the results presented above in section B.2, in the P2 condition, the best-perceived vowel is associated with a lower identification score than in the N1 condition for 9 speakers out of the 12 (S1, S2, S3, S4, S5, S7, S8, S9, S10), suggesting that compensation was poorly achieved by these speakers in this condition. This result confirms that the use of the proposed articulatory instructions can be detrimental to 4-year-old subjects. For two subjects (S11 and S12), the best perceived stimulus in P2 reached an identification score higher than in the N1 condition and greater than 80%. Thus, overall, the articulatory instructions provided in the P2 condition did not improve identification compared to P1.

### *4. Comparison of normal pre-perturbed (N1) and post-perturbed (N2) conditions*

The mean identification scores for the trials produced in the normal pre-perturbed condition N1 and the normal post-perturbed condition N2 are depicted for each subject in Figure 7. The scores obtained for the N1 condition correspond to the bars, whereas the scores obtained for the N2 condition correspond to the solid line. Comparing both values for each subject, it can be seen that, in agreement with the across-subject comparisons presented in Figure 4, a robust after-effect is observed. All speakers but three (S1, S5, S10) improved their identification scores in the N2 condition compared to the N1 condition. Two of the speakers who did not show an increase in identification score from N1 to N2 (S1 and S5) did, however, produce near-perfect scores (over 90%), suggesting a ceiling effect.

To summarize, the analysis of the production and perception data in perturbed conditions show that (1) for the majority of the subjects, the insertion of the lip-tube induces a perturbation which cannot immediately be compensated for; (2) all subjects are able to compensate for the perturbation at least once during the perturbed phase; (3) compensation strategies do not seem to be learned since no consistent improvement in production is observed from the beginning to the end of the learning phase; rather, compensation seems to be reached on a trial-and-error basis; (4) an after-effect, which suggests the existence of some kind of learning process, is observed for all subjects including those for whom our measures do not show any evidence that they were perturbed by the insertion of the lip-tube. This apparent contradiction between point (3) and point (4) will be discussed in the "General Discussion" section.

[Insert Figure 7]

### 5. *Acoustic parameters related to the target /u/*

In order to characterize the spectral parameters corresponding to the perception of the target /u/ in French and to relate those parameters to the acoustic analysis, the stimuli used in the perception experiment were plotted in various spaces determined by the spectral parameters or by combinations of these parameters in order to look for the best clustering between perceived categories. Following Savariaux *et al.* (1999) and Ménard *et al.* (2002), stimuli for which at least 50% of the listeners perceived a given vowel category and quality (referred to as dominantly perceived vowels) were plotted in spaces consisting of various combinations of F1, F2, and/or F0. Graphic representations of two acoustic-auditory spaces are provided in Figure 8. In the left panel, data were plotted in the standard F1 vs. F2 space (in Bark). It can be seen that the four categories greatly overlap. In the right panel, the F1-F0 and F2-F1 space (in Bark) is presented. These two parameters result in a much better classification of the categories. The thin solid line superimposed on the graph represents the boundary between the "not a /u/" and "bad /u/" categories. This boundary corresponds to the equation $(F2 + F1)/2 - F0 = 3.5$. Vowels for which this acoustic parameter is greater than 3.5 are perceived, in their large majority, as "not an /u/." The thick solid line in Figure 8 (right panel) corresponds to the category boundary between perceived "bad /u/" and "good /u/ or average /u/." For this boundary, the value of the parameter (F2 + F1)/2 – F0 is 3 Bark. Vowels for which this parameter is lower than 3 are identified as "good /u/" or "average /u/." Vowels for which (F2 + F1)/2 – F0 is greater than 3 Bark and lower than 3.5 Bark are perceived as "bad /u/." A discriminant analysis performed on those values with (F1-F0) and (F2-F0) as the classification parameters and the three categories "bad /u/," "not a /u/" and "average or good /u/" as the grouping factor revealed a percentage of correct classification of 85%. A similar discriminant analysis run with F1 and F2 as the classification factors and perceived category as the grouping factor resulted in a much lower percentage correct classification (70%). Thus, the addition of F0 as an acoustic parameter related to the perception of /u/ improves the classification of the data. These results are in line with previous work done on French (Savariaux *et al.*, 1999; Ménard *et al.*, 2002).

[Insert Figure 8]

*6. An interpretation of the compensation strategies in light of the perceptual data*

The results of the perceptual test led us to propose that the parameters (F2 + F1)/2 and F0, in Bark, are related to the perceived target region associated with /u/ in the acoustic-auditory space. Those parameters can therefore be used to reanalyze acoustic data and better describe speakers' compensatory strategies. Figure 9 represents each speaker's best perceived stimulus in the P1 and P2 conditions along four acoustic dimensions, in Bark: F0, F1, F2, and ((F2 + F1)/2 – F0). Mean values in the normal conditions N1 and N2 are also displayed. Two category boundaries are represented by the solid lines. Figure 9 thus represents the same data as those presented in Figure 6, but in acoustical dimensions.

[Insert Figure 9]

According to the data presented in Figure 6, in the P1 condition, seven speakers produced one perturbed /u/ with an identification score of greater than 80%: S1, S5, S6, S8, S10, S11, and S12. Those speakers can be considered as the best compensators. An examination of the acoustic values for the corresponding speakers in Figure 9 reveals various strategies. First, some speakers (S5 and S12) were able to strongly limit the impact of the perturbation with very few changes along the (F2 + F1)/2 – F0 dimension, with minimal alteration in F1 and F2 (less than 0.2 Bark). Thus, these speakers were able to achieve a complete compensation for the perturbation for at least one /u/ among the 20 trials by producing almost identical F1 and F2 values as in N1. For some other speakers (S8, S10, and S11), F2 was increased by values ranging from 0.7 Bark to 1.2 Bark in P1 compared to N1, but F1 was decreased by values ranging from 0.5 Bark to 1.7 Bark, thus maintaining the (F2 + F1)/2 – F0 complex below 3.5 Bark. Four other subjects increased F0 values, hypothetically to counterbalance the increase of F2. The increase in F0 varies from 0.3 Bark to 0.6 Bark (see S11, in Figure 9, upper left panel), values in the range of those found by Savariaux *et al.* (1999). Although the interpretation of an active control of F1, F2, and F0 to maintain a low value of the (F2 + F1)/2 – F0 parameter is highly speculative, the fact that compensation can sometimes be observed in the (F2 + F1)/2 – F0 dimension but not in the F1, F2, and F0 dimensions separately reveals that the degrees of freedom required to produce an acoustic-auditory target are known and controlled early in life.

## IV. GENERAL DISCUSSION

### A. Compensatory mechanisms in children

The results presented so far demonstrated that compensatory abilities in 4-year-old French children are speaker-dependent, a pattern also found in adults (Savariaux *et al.*, 1995; 1999). Perceptual data showed that upon insertion of the lip-tube, in the first perturbed trial, two speakers (S6 and S12) were immediately able to compensate in the P1 condition. Indeed, those speakers produced /u/ with an identification score greater than, or equal to their mean identification scores in the N1 condition. Two speakers (S6 and S11) also achieved immediate compensation in the first trial in the P2 condition, as revealed by the identification scores. The remaining speakers, however, did not demonstrate compensatory abilities at the first perturbed trial, based on the variation in identification scores in the perturbed conditions compared to those in the normal pre-perturbed condition N1. Interestingly, no systematic improvement was observed during the perturbed phase, but in the P1 condition, all speakers have the articulatory skills to produce at least one good compensation (see Figure 6). This shows (1) that all the subjects were able to compensate for the perturbation, and (2) that the motor control of speech is still too immature in 4-year-old children to allow generalization and learning in such a short time.

Turning now to the P2 condition, we found that only one speaker increased his identification score over the 20 trials (but to a value of 8%). Thus, for almost all speakers, the articulatory instructions provided in this condition were detrimental to the creation of successful compensatory strategies. One could object that speakers may not have understood the instructions related to the P2 condition. However, this is very unlikely since children were reminded during this perturbed phase by the experimenters that they had to go from /o/ and reach /u/.

It should be noted that none of the speakers demonstrated a real learning mechanism during the perturbed trials (either P1 or P2). Indeed, no significant correlations between trial number and identification scores were obtained, suggesting that the compensation strategies were not obtained by a gradual error correction mechanism operating on articulatory maneuvers. Rather, our 4-year-old subjects seem to have produced a good target in the presence of the lip-tube on a trial-and-error basis, as revealed by the larger variability (measured by standard deviation) found in P1 and P2 compared to N1 and N2. Even though most of the children produced alterations along the acoustic dimensions that can be interpreted as strategies allowing them to achieve better compensation in the acoustic-auditory space, they could not store those strategies. This can be explained by the immaturity of their internal model of the articulatory-acoustic relations. This pattern contrasts with the adult data presented in Savariaux *et al.* (1995, 1999), in which speakers showed evidence of gradual improvement in the articulatory-acoustic domain. In this respect, it appears that 4-year-old children have a limited knowledge of the relationships between articulatory maneuvers and their acoustic consequences. Nevertheless, a robust after-effect is observed in P1 and P2 conditions, even for the subjects who did not compensate. This contradiction deserves further analysis.

### B. The detrimental effect of articulatory guidance

Identification scores in the P1 condition suggests that, for almost all subjects, articulatory changes were produced in order to reach the acoustic-auditory target. Changes in tongue shape and position, however, were likely local and did not involve movements as large as those required to fully compensate. As discussed in section II.A.2, simulations with VLAM have revealed that /u/'s formant frequencies are affiliated to Helmholtz resonators. Thus, local changes in constriction area and/or constriction length may have been done in order to modify formant frequencies while preserving as much as possible tongue position related to the original /u/. Thus, when lip area is increased, small changes in tongue position can alter formant values and enhance identification scores, even though those strategies are not optimal. For some speakers (because of the morphology of their vocal tract, for instance), however, this task may have been more difficult because of the changes of articulatory position these compensatory maneuvers required.

Another compensatory strategy predicted by our simulations with VLAM involves a large displacement of the tongue to a position closer to that in /o/. The articulatory cues provided in the P2 condition were intended to guide the speakers to this latter configuration. However, in the P2 condition, it can be hypothesized that the compensatory maneuvers were less successful because of the large distance between the produced somatosensory target and the intended somatosensory target (induced by the articulatory instructions concerning tongue position), the articulatory alterations induced by the instructions being too far from the somatosensory target associated with /u/. Thus, large changes in lingual articulatory configurations were not produced. This hypothesis is currently tested with articulatory measurements.

### C. The nature of the speech task for 4-year-old children

The fact that over the course of the 20 trials, each subject has achieved a good compensation in the spectral domain associated with good perception scores supports the hypothesis that these subjects were trying to reach a target in the auditory domain. The observation that none of the subjects kept producing the vocal tract configuration associated with a good auditory product certainly suggests that the auditory objective could have been not satisfactory *per se*. In addition, the inability of all subjects to achieve compensation in the perceptual domain in P2 condition suggests that somatosensory aspects contribute also to the specification of the task. In this context, one can not completely discard the possibility that the discomfort associated with an unusual speaking condition could have been responsible for the larger intra-speaker variability observed in P1 and P2 conditions, simply because it forced the subjects to move their articulators around the canonical vocal tract configuration without any specific auditory goal. However, in such conditions the probability for every speaker to reach by coincidence the right auditory target over the course of a reduced number of trials would have been quite low. This is why, in line of Savariaux et al.'s (1995, 1999) conclusions for adults, we strongly favour the hypothesis of an auditory target for 4-year-old children as well. In this context, the absence of systematic and continuous improvement over the course of the 20 trials in the P1 condition is explained by the fact that children would not integrate the auditory feedback in a form that can be processed in terms of learning because of immature internal models. This result points to the very distinct nature of the speech representations in 4-year-old children and adults. In the present study, children did not demonstrate a good knowledge of their articulatory capabilities in relation to their impact on acoustics. A closed-loop correction of articulatory positions on the basis of the minimization of the difference between the intended auditory feedback and the auditory feedback actually produced did not occur. This pattern is very different from the one obtained in adults by Savariaux *et al.* (1995, 1999). In the latter study, even though almost all speakers failed to achieve complete compensation by the end of the perturbed condition, all of them had produced some articulatory-acoustic maneuvers leading to a minimization of the discrepancy between intended and produced /u/.

The presence of the robust after-effect is nonetheless intriguing. How could speakers use different strategies and improve their identification scores after removal of the tube if no learning mechanism had occurred in the perturbed conditions? We hypothesize that children establish associative links between multisensory representations (phonemic) and articulatory maneuvers, but that those links are not yet internalized in the form of an internal model of speech control. Children aim to produce a good acoustic-auditory target /u/ during the perturbed trials. After removal of the tube, their goal is still to produce a good target. In all likelihood, then, the observed after-effect reflects the speakers' efforts to produce the canonical values of /u/ and increase their identification scores, rather than the maintenance of any compensatory maneuvers they used earlier. Further studies designed to investigate tongue shape and position in such perturbed conditions are currently in progress, with the hope that they will shed more light on this issue.

**ACKNOWLEDGMENTS**

This work was supported by the Social Sciences and Humanities Research Council of Canada (SSHRC), the Natural Sciences and Engineering Research Council of Canada (NSERC) and the Fonds Québécois de Recherche sur la Société et la Culture (FQRSC). The authors would like to thank David H. McFarland for insightful discussions. We are grateful to Zofia Laubitz for copy-editing the paper.

---

[i] The Bonferonni correction, which would have resulted in adapting the probability level to 0.0125, was not applied here, in order to follow as strictly as possible the method previously used in Savariaux *et al.*(1995, 1999) and Baum and Katz (1988).

TABLE 1: Mean F1 values (in Hertz) for each subject, in the four experimental conditions and at two measurement points (onset and midpoint). Standard deviations are presented in square brackets. Percentage variation relative to the N1 condition is presented in parentheses.

|  | Onset | | | | | Midpoint | | | |
|---|---|---|---|---|---|---|---|---|---|
|  | N1 | P1 | P2 | N2 | | N1 | P1 | P2 | N2 |
| S1 P1-P2 | 434 [41] | 395 [15] (−9%) | 571 [32] (+32%) | 410 [11] (−5%) | | 396 [21] | 414 [15] (+5%) | 444 [25] (+12%) | 423 [14] (+7%) |
| S2 P1-P2 | 443 [58] | 553 [108] (+25%) | 573 [63] (+29%) | 356 [0] (−20%) | | 394 [37] | 513 [77] (+30%) | 484 [52] (+23%) | 348 [18] (−12%) |
| S3 P1-P2 | 472 [46] | 556 [98] (+18%) | 618 [78] (+31%) | 439 [40] (−7%) | | 444 [35] | 439 [34] (−1%) | 520 [66] (+17%) | 420 [46] (−5%) |

| | | | | | | | | |
|---|---|---|---|---|---|---|---|---|
| S4 P2-P1 | 518 [74] | 611 [71] (+18%) | 681 [90] (+31%) | 491 [55] (−5%) | 490 [110] | 566 [70] (+15%) | 623 [82] (+27%) | 427 [90] (−13%) |
| S5 P2-P1 | 465 [131] | 581 [100] (+25%) | 638 [86] (+37%) | 396 [24] (−15%) | 387 [10] | 401 [24] (+4%) | 561 [69] (+45%) | 371 [32] (−4%) |
| S6 P2-P1 | 449 [55] | 629 [194] (+40%) | 619 [96] (+38%) | 508 [63] (+13%) | 445 [28] | 506 [111] (+14%) | 523 [114] (+17%) | 413 [25] (−7%) |
| S7 P1-P2 | 360 [15] | 530 [128] (+47%) | 540 [151] (+50%) | 407 [31] (+13%) | 534 [65] | 476 [77] (−11%) | 550 [34] (+3%) | 426 [70] (−20%) |
| S8 P1-P2 | 447 [57] | 639 [111] (+43%) | 633 [86] (+42%) | 403 [27] (−10%) | 414 [42] | 527 [70] (+27%) | 502 [36] (+21%) | 355 [27] (−14%) |
| S9 P1-P2 | 362 [83] | 556 [124] (+54%) | 650 [136] (+80%) | 374 [15] (+3%) | 491 [104] | 549 [55] (+12%) | 533 [56] (+9%) | 390 [79] (−20%) |
| S10 P2-P1 | 472 [38] | 587 [153] (+24%) | 640 [190] (+36%) | 401 [65] (−15%) | 501 [55] | 468 [46] (−7%) | 399 [95] (−20%) | 415 [43] (−17%) |
| S11 P2-P1 | 419 [28] | 556 [70] (+33%) | 588 [69] (+40%) | 429 [48] (+2%) | 571 [41] | 573 [80] (0%) | 551 [78] (−3%) | 505 [36] (−12%) |
| S12 P2-P1 | 427 [54] | 555 [132] (+30%) | 546 [71] (+28%) | 411 [32] (−4%) | 382 [17] | 425 [68] (+11%) | 379 [47] (−1%) | 353 [25] (−8%) |

TABLE 2: Mean F2 values (in Hertz) for each subject, in the four experimental conditions and at two measurement points (onset and midpoint). Standard deviations are presented in square brackets. Percentage variation relative to the N1 condition is presented in parentheses.

| | Onset | | | | | Midpoint | | | |
|---|---|---|---|---|---|---|---|---|---|
| | N1 | P1 | P2 | N2 | | N1 | P1 | P2 | N2 |
| S1 P1-P2 | 951 [54] | 1118 [70] (+18%) | 1313 [78] (+38%) | 904 [51] (−5%) | | 1075 [39] | 1175 [48] (+9%) | 1292 [115] (+20%) | 991 [22] (−8%) |
| S2 P1-P2 | 808 [61] | 1390 [232] (+72%) | 1348 [141] (+67%) | 695 [24] (−14%) | | 839 [89] | 1255 [152] (+50%) | 1240 [74] (+48%) | 726 [71] (−13%) |
| S3 P1-P2 | 944 [85] | 1237 [152] (+31%) | 1265 [139] (+34%) | 703 [113] (−26%) | | 982 [127] | 1105 [113] (+12%) | 1123 [85] (+14%) | 709 [49] (−28%) |
| S4 P2-P1 | 1005 [112] | 1350 [196] (+34%) | 1244 [235] (+24%) | 888 [101] (−12%) | | 1024 [171] | 1110 [55] (+8%) | 1099 [66] (+7%) | 929 [83] (−9%) |
| S5 P2-P1 | 997 [26] | 1250 [154] (+25%) | 1185 [179] (+19%) | 817 [50] (−18%) | | 1035 [64] | 1074 [47] (+4%) | 1105 [121] (+7%) | 809 [87] (−22%) |
| S6 P2-P1 | 859 [115] | 1434 [216] (+67%) | 1350 [154] (+57%) | 1009 [106] (+17%) | | 971 [35] | 1176 [147] (+21%) | 1212 [162] (+25%) | 1006 [96] (+4%) |
| S7 P1-P2 | 752 [47] | 1495 [485] (+99%) | 1176 [217] (+56%) | 693 [39] (−8%) | | 1204 [134] | 1110 [136] (−8%) | 1105 [74] (−8%) | 775 [100] (−36%) |
| S8 P1-P2 | 790 [98] | 1362 [390] (+72%) | 1251 [280] (+58%) | 739 [60] (−6%) | | 838 [38] | 1147 [137] (+37%) | 1043 [75] (+25%) | 814 [105] (−3%) |
| S9 P1-P2 | 691 [84] | 1220 [255] (+77%) | 1278 [282] (+85%) | 701 [50] (+1%) | | 884 [189] | 1124 [121] (+27%) | 1041 [91] (+18%) | 720 [92] (−19%) |
| S10 P2-P1 | 794 [84] | 1352 [445] (+70%) | 1379 [547] (+74%) | 788 [119] (−1%) | | 857 [103] | 981 [91] (+15%) | 1005 [108] (+17%) | 792 [65] (−8%) |
| S11 P2-P1 | 820 [66] | 1246 [204] (+52%) | 1215 [345] (+48%) | 725 [42] (−12%) | | 1050 [85] | 1164 [100] (+11%) | 1037 [71] (−1%) | 872 [97] (−17%) |
| S12 P2-P1 | 929 [69] | 1327 [429] (+43%) | 1160 [332] (+25%) | 810 [45] (−13%) | | 917 [48] | 956 [127] (+4%) | 915 [73] (0%) | 827 [38] (−10%) |

TABLE 3: Mean F0 values (in Hertz) for each subject, in the four experimental conditions and at two measurement points (onset and midpoint). Standard deviations are presented in square brackets. Percentage variation relative to the N1 condition is presented in parentheses.

|  | Onset | | | | | Midpoint | | | |
|---|---|---|---|---|---|---|---|---|---|
|  | N1 | P1 | P2 | N2 | | N1 | P1 | P2 | N2 |
| S1 P1-P2 | 323 [46] | 362 [13] (+12%) | 313 [60] (−3%) | 378 [8] (+17%) | | 364 [16] | 384 [19] (+5%) | 428 [34] (+18%) | 416 [13] (+14%) |
| S2 P1-P2 | 262 [34] | 317 [29] (+21%) | 256 [35] (−2%) | 292 [27] (+12%) | | 308 [9] | 323 [12] (+5%) | 267 [16] (−13%) | 316 [5] (+2%) |
| S3 P1-P2 | 277 [31] | 260 [20] (−6%) | 244 [25] (−12%) | 231 [10] (−17%) | | 236 [3] | 240 [7] (+2%) | 241 [14] (+2%) | 238 [5] (+1%) |
| S4 P2-P1 | 286 [46] | 298 [38] (+4%) | 285 [41] (0%) | 260 [28] (−9%) | | 294 [15] | 296 [6] (+1%) | 283 [12] (−4%) | 295 [9] (+1%) |
| S5 P2-P1 | 366 [22] | 356 [33] (−3%) | 323 [19] (−12%) | 335 [28] (−8%) | | 374 [10] | 357 [12] (−5%) | 315 [15] (−16%) | 335 [16] (−11%) |
| S6 P2-P1 | 386 [14] | 371 [14] (−4%) | 374 [16] (−3%) | 352 [59] (−9%) | | 348 [11] | 364 [14] (+5%) | 356 [19] (+2%) | 356 [19] (+2%) |
| S7 P1-P2 | 313 [19] | 301 [21] (−4%) | 295 [36] (−6%) | 304 [45] (−3%) | | 295 [7] | 305 [10] (+3%) | 288 [8] (−2%) | 290 [13] (−2%) |
| S8 P1-P2 | 322 [27] | 302 [24] (−6%) | 285 [33] (−12%) | 316 [17] (−2%) | | 314 [4] | 315 [14] (0%) | 268 [10] (−15%) | 303 [11] (−4%) |
| S9 P1-P2 | 301 [62] | 312 [41] (+4%) | 298 [37] (−1%) | 324 [26] (+8%) | | 311 [31] | 321 [22] (+3%) | 271 [21] (−13%) | 297 [17] (−5%) |
| S10 P2-P1 | 255 [15] | 252 [19] (−1%) | 237 [31] (−7%) | 232 [15] (−9%) | | 279 [20] | 250 [17] (−10%) | 215 [13] (−23%) | 244 [20] (−13%) |
| S11 P2-P1 | 342 [26] | 319 [53] (−7%) | 297 [75] (−13%) | 291 [25] (−15%) | | 316 [6] | 330 [18] (+5%) | 330 [16] (+4%) | 287 [7] (−9%) |
| S12 P2-P1 | 340 [37] | 328 [46] (−4%) | 320 [32] (−6%) | 341 [33] (0%) | | 312 [16] | 335 [26] (+8%) | 313 [15] (0%) | 314 [19] (+1%) |

TABLE 4: Mean percentage of produced [u] perceived as [u] for each subject, in the four experimental conditions. Percentage variation relative to the N1 condition is presented in parentheses.

|  | N1 | P1 | P2 | N2 |
|---|---|---|---|---|
| S1 $_{P1-P2}$ | 91 | 79 (–13%) | 86 (–5%) | **91 (0%)** |
| S2 $_{P1-P2}$ | 73 | 55 (–25%) | 15 (–80%) | 97 (+33%) |
| S3 $_{P1-P2}$ | 60 | 23 (–61%) | 1 (–98%) | 77 (+29%) |
| S4 $_{P2-P1}$ | 47 | 19 (–60%) | 1 (–99%) | **100 (+114%)** |
| S5 $_{P2-P1}$ | 96 | 87 (–9%) | 18 (–81%) | **93 (–3%)** |
| S6 $_{P2-P1}$ | 73 | 48 (–35%) | 42 (–43%) | **92 (+25%)** |
| S7 $_{P1-P2}$ | 61 | 45 (–26%) | 3 (–96%) | **79 (+28%)** |
| S8 $_{P1-P2}$ | 88 | 34 (–61%) | 7 (–92%) | **97 (+11%)** |
| S9 $_{P1-P2}$ | 53 | 24 (–55%) | 11 (–79%) | 80 (+50%) |
| S10 $_{P2-P1}$ | 73 | 36 (–51%) | 8 (–89%) | 69 (–5%) |
| S11 $_{P2-P1}$ | 43 | 35 (–17%) | 36 (–16%) | **80 (+88%)** |
| S12 $_{P2-P1}$ | 61 | 68 (+11%) | 59 (–3%) | 100 (+63%) |

Figure captions :

FIGURE 1 : Predicted values of F1 and F2 in the normal condition (Normal), with the lip-tube inserted but with no compensation (Pert), and with the lip-tube inserted and compensation strategies (Pert-comp : backward and downward movement of the tongue). Values are in percentage variation compared to the normal condition.

FIGURE 2 : Mean and standard deviation values of F1 (left panel) and F2 (right panel) across subjects, in the four experimental conditions (N1, P1, P2, N2). Values measured at vowel onset correspond to the solid line, and values measured at vowel midpoint are depicted by the dashed line. All values in Hertz.

FIGURE 3 : Mean and standard deviation values of F1 (left panel) and F2 (right panel) across subjects, in the four experimental conditions (N1, P1, P2, N2), for the first (solid line) and last trials (dashed line). All values in Hertz.

FIGURE 4 : Mean and standard deviation values of identification scores across subjects, in the four experimental conditions (N1, P1, P2, N2).

FIGURE 5 : Identification scores of first (solid line) and last (dashed line) trials in P1 (left panel) and P2 (right panel), for each subject. Mean scores in the N1 condition are provided and correspond to the bars.

FIGURE 6 : Identification scores of the best perceived stimulus in P1 (solid line) and P2 (dashed line), for each subject. Mean scores in the N1 condition are provided and correspond to the bars.

FIGURE 7 : Mean identification scores in N1 and N2 conditions for each subject. Values for N1 correspond to the bars; values for N2 correspond to the solid line.

FIGURE 8 : Dominantly perceived stimuli in the F1 vs. F2 space (left panel) and in the F2-F1 vs F1-F0 space (right panel). All values in bark. Circles correspond to stimuli rated « not an /u/ », crosses correspond to stimuli rated « bad », plusses correspond to stimuli rated « good », triangles correspond to stimuli rated « average ». The thin solid line (right panel) corresponds to the category boundary between perceived "not a /u/" and "bad /u/", for which $(F2+F1)/2-F0=3.5$ Bark. The thick solid line (right panel) corresponds to the category boundary between perceived "bad /u/" and "perceived average or good /u/", for which $(F2+F1)/2-F0=3$ Bark.

FIGURE 9 : Values of F0 (upper left panel), F1 (upper right panel), F2 (lower left panel), and $(F2+F1)/2-F0$ (lower right panel) of the best perceived stimulus in each of the perturbed conditions P1 and P2, for each speaker. Mean values in N1 and N2 conditions are also displayed. All values in Bark.

FIGURE 1 :

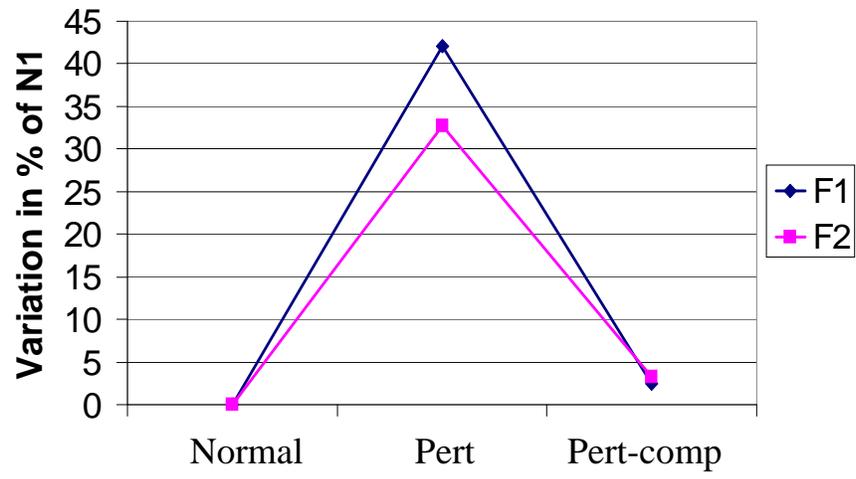

Figure 2 :

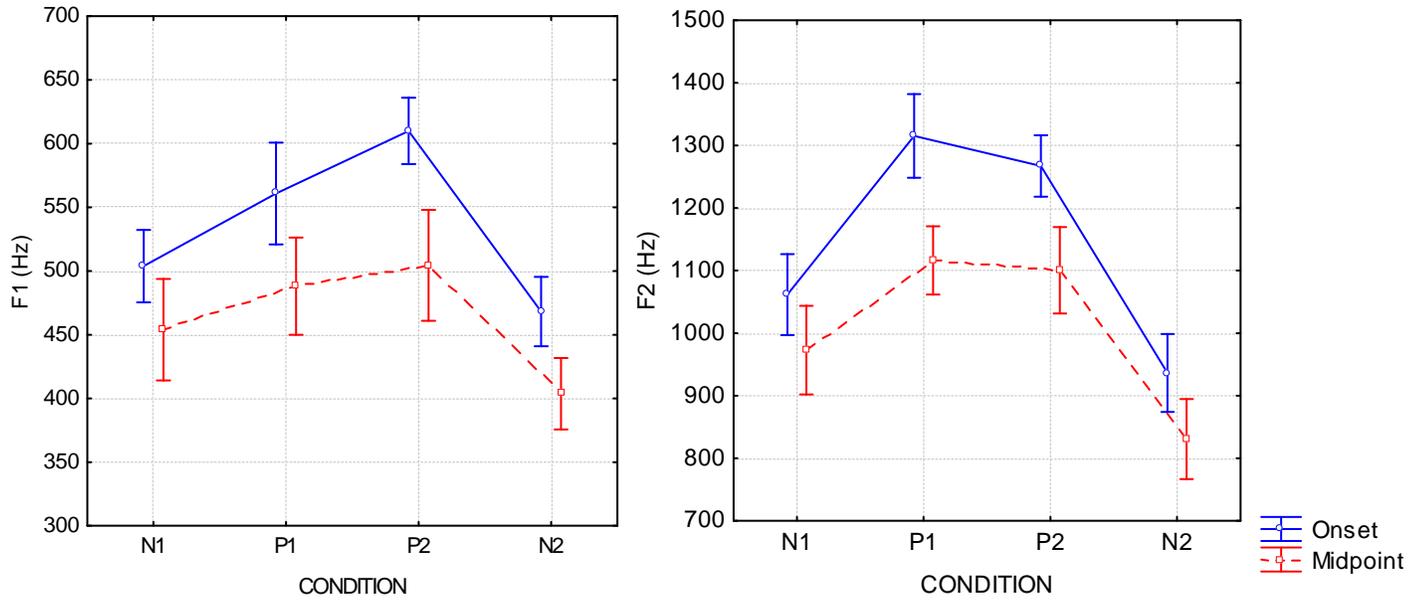

Figure 3:

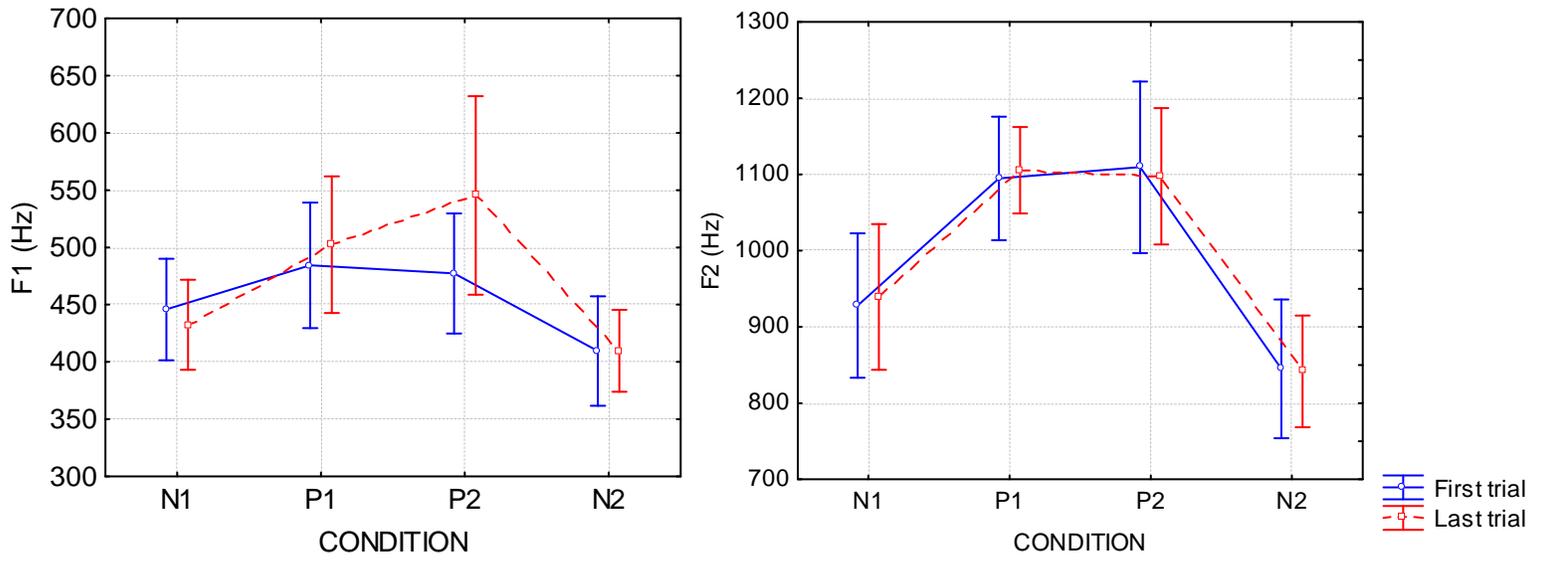

Figure 4:

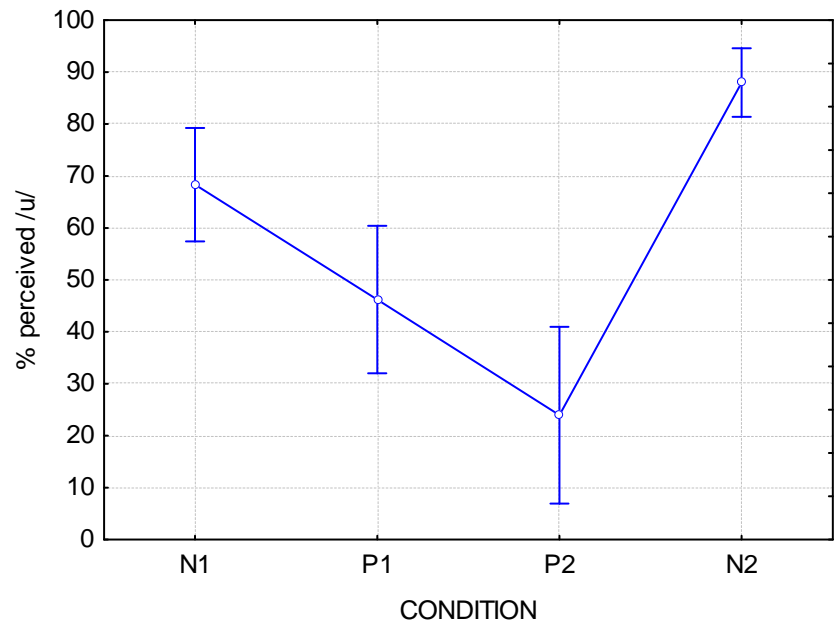

Figure 5 :

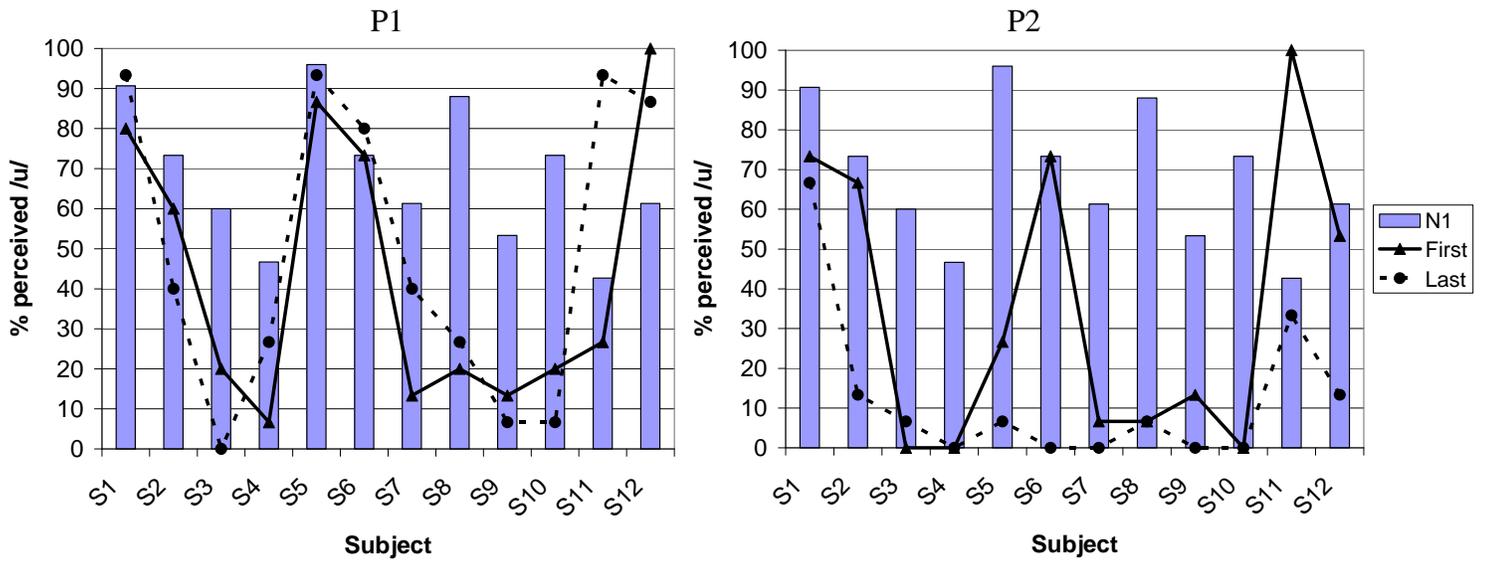

Figure 6 :

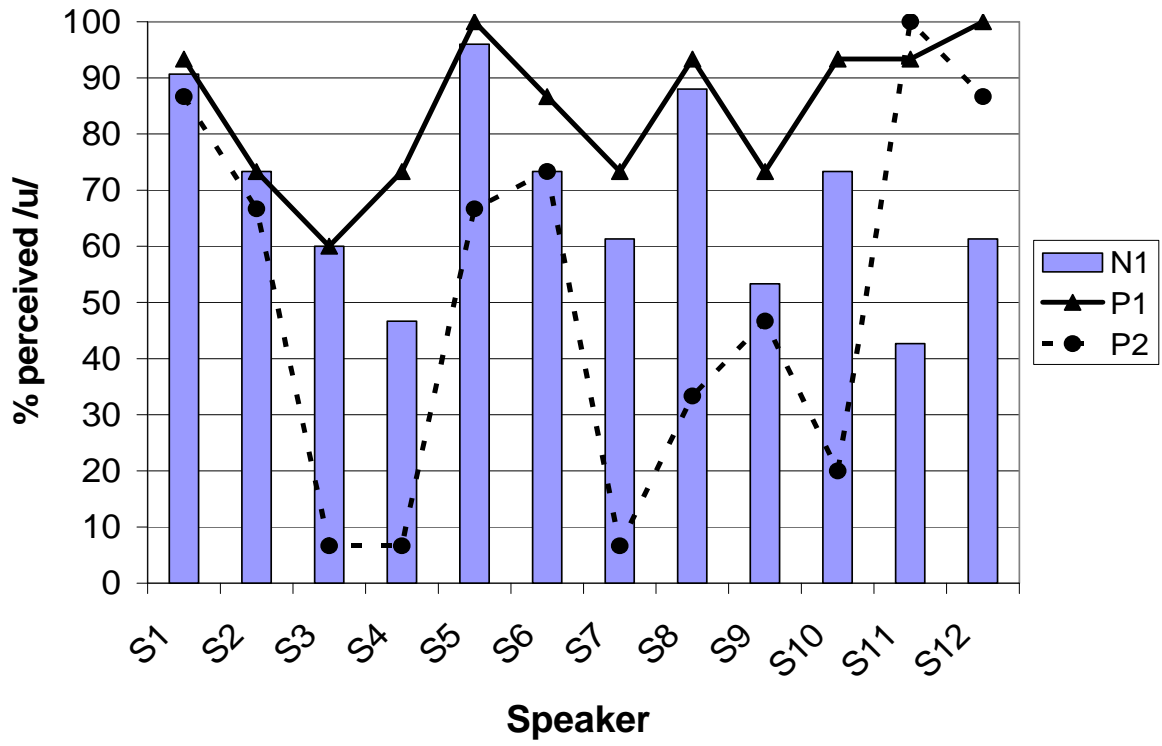

Figure 7 :

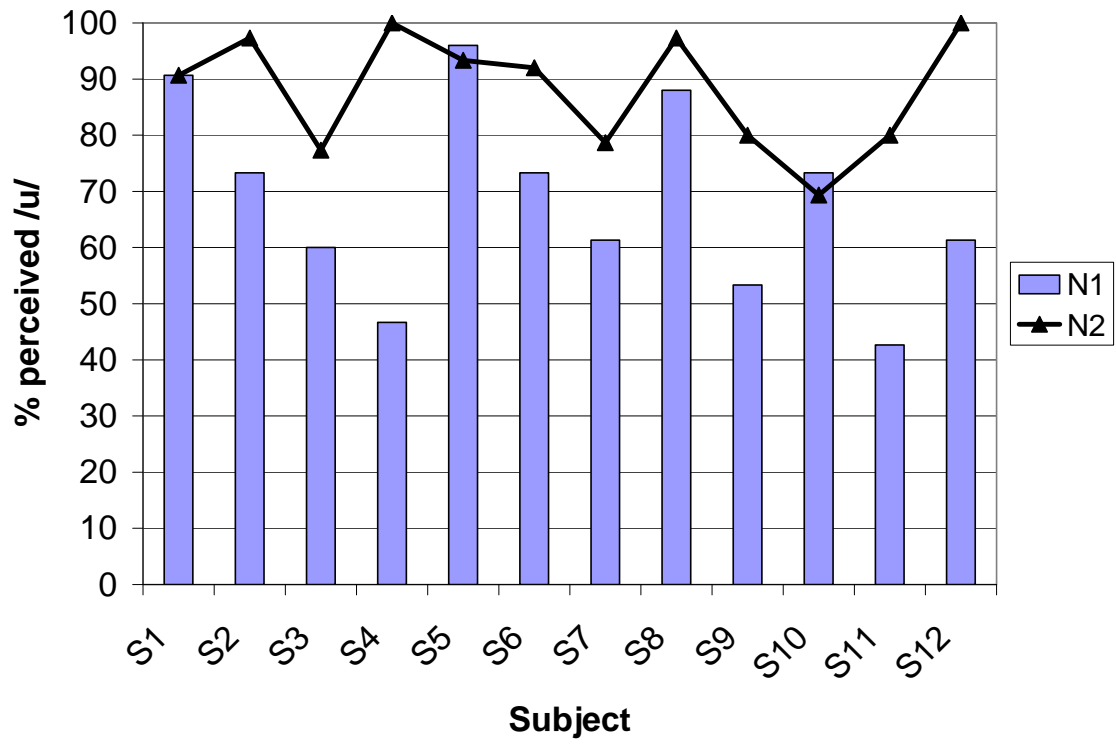

Figure 8 :

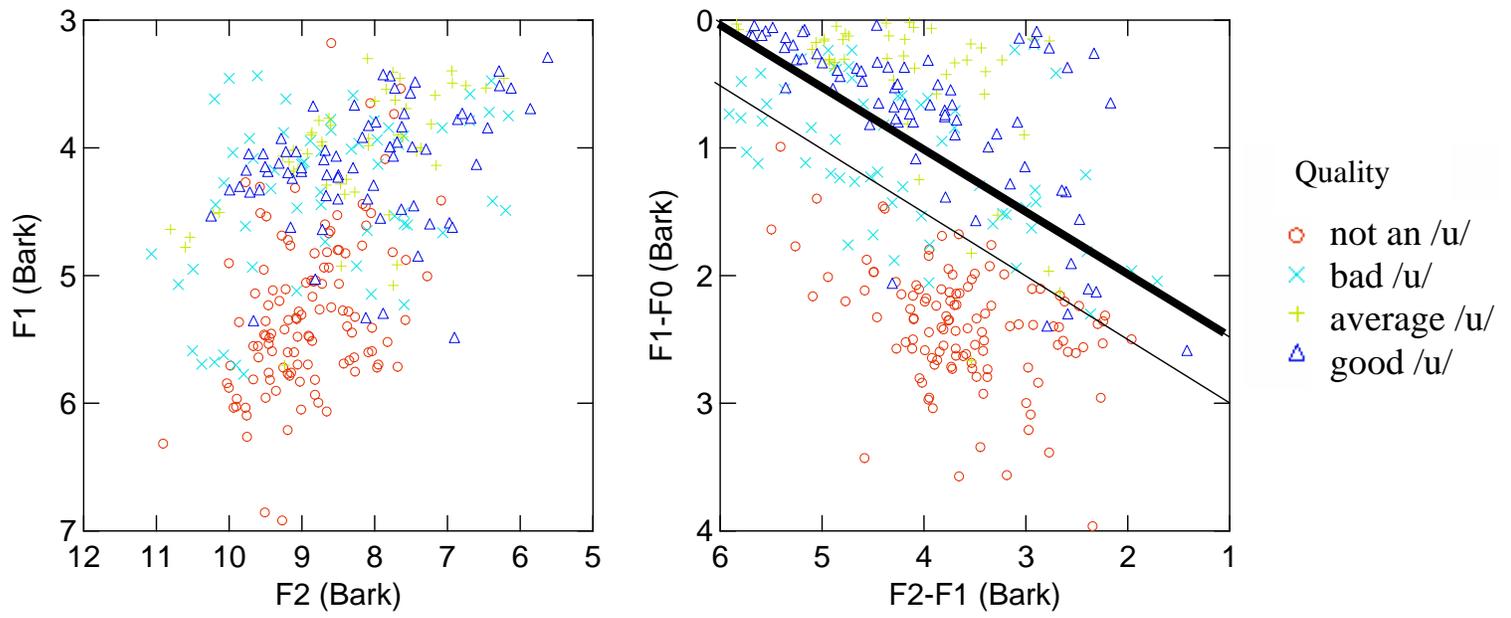

Figure 9 :

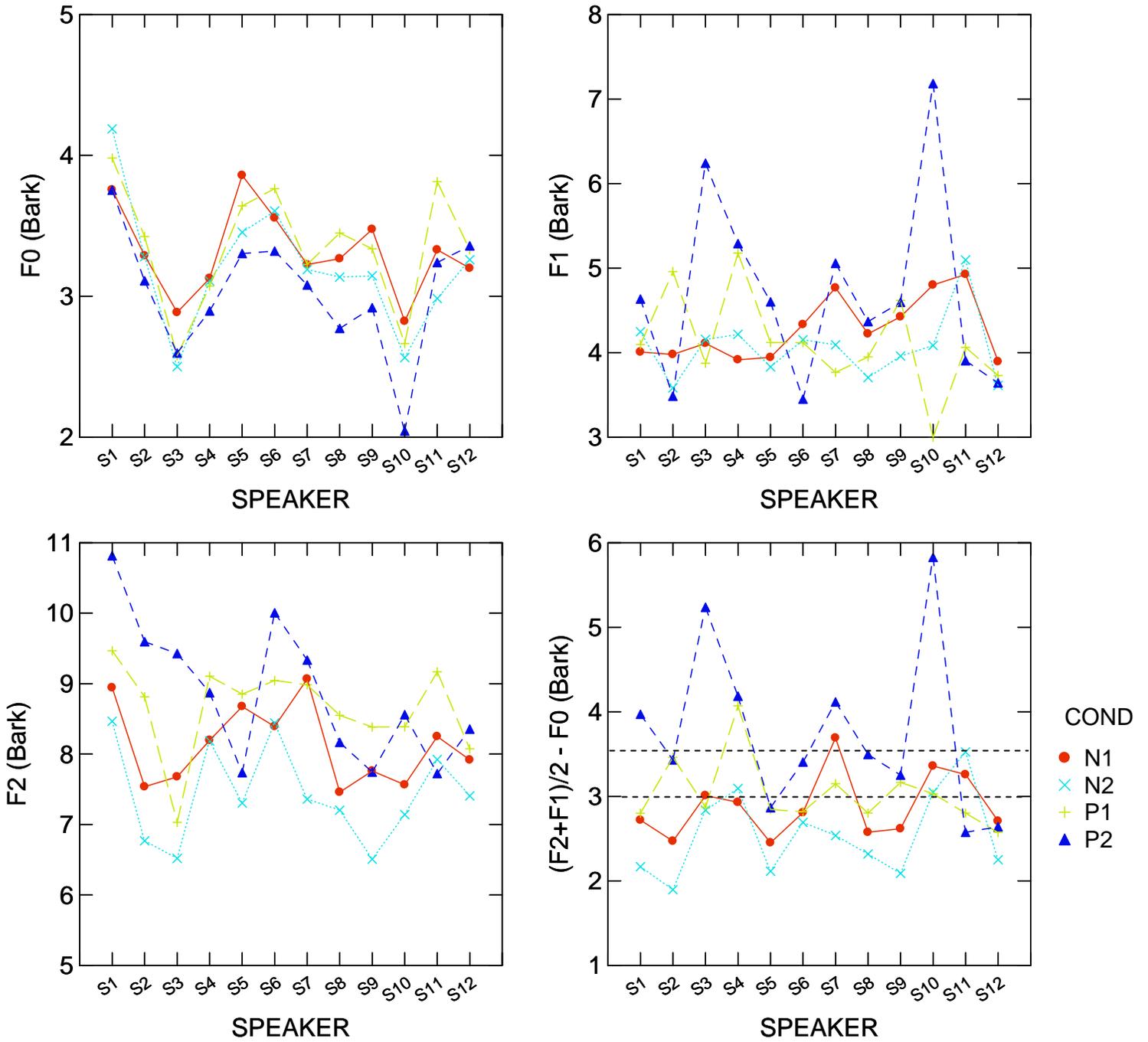